\documentclass{article}
\usepackage{amsfonts}

\usepackage{amsmath}
\usepackage{graphicx}

\newtheorem{theorem}{Theorem}

\newtheorem{definition}[theorem]{Definition}

\input{tcilatex}

\begin{document}

\title{Filtering of Wide Sense Stationary \\
Quantum Stochastic Processes}
\author{John Gough \\
%EndAName
Institute for Mathematical and Physical Sciences,\\
University of Wales, Aberystwyth,\\
Ceredigion, SY23 3BZ, Wales}
\date{}
\maketitle

\begin{abstract}
We introduce a concept of a quantum wide sense stationary process taking
values in a C*-algebra and expected in a sub-algebra. The power spectrum of
such a process is defined, in analogy to classical theory, as a positive
measure on frequency space taking values in the expected algebra. The notion
of linear quantum filters is introduced as some simple examples mentioned.
\end{abstract}

\bigskip

\section{Spectral Analysis of WSS Quantum Processes}

In the following we consider processes depending on a continuous time
variable $t\in \mathbb{R}$: the frequency domain will be denoted as $\mathbb{%
\hat{R}}$ for distinction. Therefore we have the Fourier transform
conventions $\hat{f}\left( \nu \right) =\int_{\mathbb{R}}e^{2\pi i\,t\nu
}\,f\left( t\right) dt$, $f\left( t\right) =\int_{\mathbb{\hat{R}}}e^{-2\pi
i\,t\nu }\,\hat{f}\left( \nu \right) d\nu $.

\subsection{Classical Theory of WSS Processes}

Let $\frak{h}=L^{2}\left( \Omega ,\mathcal{F},\mathbb{P}\right) $ be the
Hilbert space of second-order random variables on a probability space $%
\left( \Omega ,\mathcal{F},\mathbb{P}\right) $. A second order process $%
X=\left( X_{t}\right) _{t}$ is a stochastic process with each $X_{t}\in 
\frak{h}$. (That is, $X$\ corresponds to a parameterized curve in the
Hilbert space $\frak{h}$. The linear manifold in $\frak{h}$\ spanned by the
family $\left\{ X_{t}:t\in \mathbb{R}\right\} $ is denote as $\frak{l}_{X}$\
and its closure $\frak{h}_{X}$ is a Hilbert subspace of $\frak{h}$. The mean
function is $\mu _{X}\left( t\right) :=\mathbb{E}\left[ X_{t}\right] $\ and
the covariance function is $C_{X}\left( s,t\right) :=\mathbb{E}\left[
X_{s}^{\ast }X_{t}\right] -\mu _{X}\left( t\right) \mu _{X}\left( s\right)
\equiv \langle X_{s},X_{t}\rangle _{\frak{h}}-\mu _{X}\left( t\right) \mu
_{X}\left( s\right) $. If $\mu _{X}$ is independent of $t$ and $C_{X}\left(
s,t\right) $ depends only on $t-s$ then the process is termed wide sense
stationary (WSS).

If we are given a mean-zero WSS process $X$ then we write its covariance
function as $C_{X}\left( \tau \right) \equiv C_{X}\left( t,t+\tau \right) $.
The covariance is semi-positive definite ( $\sum_{j,k=1}^{n}z_{j}^{\ast
}C_{X}\left( t_{j},t_{k}\right) z_{k}\geq 0$ or all finite integers $n$, for
arbitrary times $t_{1},\dots ,t_{n}\in \mathbb{R}$ and arbitrary complex
numbers $z_{1},\dots ,z_{n}$) and by Bochner's theorem (Herglotz' theorem
for discrete time parameter) it can therefore be written as a
Fourier-Stieltjes transform \cite{B}\cite{GS} 
\begin{equation*}
C_{X}\left( t\right) =\int_{\mathbb{\hat{R}}}e^{2\pi i\,t\nu }\,dS_{X}\left(
\nu \right)
\end{equation*}
where $S_{X}\left( \nu \right) $ is a non-decreasing right-continuous on $%
\mathbb{\hat{R}}$ with $\lim_{\nu \rightarrow -\infty }S_{X}\left( \nu
\right) =0$ and $\lim_{\nu \rightarrow +\infty }S_{X}\left( \nu \right)
=C_{X}\left( 0\right) $. $S_{X}$ is called the spectral function for $X$.

\bigskip

Let $\frak{E}$\ be the linear manifold spanned by the harmonic functions $%
e_{t}\left( \nu \right) =\exp \left( 2\pi i\,t\nu \right) $ on $\mathbb{\hat{%
R}} $.\ We set $\frak{f}_{X}=L^{2}\left( \mathbb{\hat{R}},dS_{X}\right) $.
Evidently $\frak{E}$\ is a subset of $\frak{f}_{X}$ which is dense in the $%
\left\| .\right\| _{\frak{f}_{X}}$-topology.

\bigskip

\begin{theorem}
Let $X$ be a mean-zero WSS process then there exists a linear isometric
isomorphism $\hat{X}:\frak{f}_{X}\mapsto \frak{h}_{X}$.
\end{theorem}

The proof is constructive: define $\hat{X}:\frak{E}\mapsto \frak{h}_{X}$
first of all by setting $\hat{X}\left( e_{t}\right) \equiv X_{t}$ and
extending by linearity. It follows that $\hat{X}$ is an isometry, since 
\begin{eqnarray*}
\left\langle \hat{X}\left( e_{s}\right) |\hat{X}\left( e_{t}\right)
\right\rangle _{\frak{h}} &=&C_{X}\left( t-s\right) \\
&=&\int_{\mathbb{\hat{R}}}e_{s}\left( \nu \right) ^{\ast }e_{t}\left( \nu
\right) \,dS_{X}\left( \nu \right) \\
&=&\left\langle e_{s}|e_{t}\right\rangle _{\frak{f}_{X}}.
\end{eqnarray*}
Next given any $Z\in \frak{h}_{X}$ there will be a sequence $\left(
Z_{n}\right) _{n}$ converging to $Z$ in norm such that each $Z_{n}\equiv 
\hat{X}\left( f_{n}\right) $ for some $f_{n}\in \frak{E}$. The sequence of
function $\left( f_{n}\right) _{n}$ will have a well-defined limit $f$\ in $%
\frak{f}_{X}$ depending only on $Z$. We therefore set $Z=\hat{X}\left(
f\right) $. Existence and uniqueness follows from the completeness of the
Hilbert spaces $\frak{f}_{X}$ and $\frak{h}_{X}$.

\bigskip

The map $\hat{X}:\frak{f}_{X}\mapsto \frak{h}_{X}$ induces a $\frak{h}_{X}$%
-valued measure on the Borel sets of $\mathbb{\hat{R}}$, again denoted as $%
\hat{X}\left[ .\right] $, according to $\hat{X}\left[ F\right] \equiv \hat{X}%
\left( 1_{F}\right) $ where $1_{F}$ is the characteristic function for the
subset $F$. In this way we can write 
\begin{equation*}
\hat{X}\left( f\right) \equiv \int_{\mathbb{\hat{R}}}f\left( \nu \right) \,%
\hat{X}\left[ d\nu \right]
\end{equation*}
and in particular $X_{t}=\hat{X}\left( e_{t}\right) \equiv \int_{\mathbb{%
\hat{R}}}e^{2\pi i\,t\nu }\,\hat{X}\left[ d\nu \right] $. (Typically a
Fourier transform $\hat{X}_{\nu }$ may only exist in some more singular
sense.)

\paragraph{Remark (1)}

The set $\frak{E}$ in fact forms a commutative *-algebra with unit $e_{0}$.

\paragraph{Remark (2)}

If $X$ and $Y$ are orthogonal (independent!) mean-zero WSS processes then $%
dS_{\alpha X+\beta Y}=|\alpha |^{2}dS_{X}+|\beta |^{2}dS_{Y}$.

\paragraph{Remark (3)}

For each mean-zero WSS process $X$ we construct a kernel $K_{X}:\frak{f}%
_{X}\times \frak{f}_{X}\mapsto C$ according to 
\begin{equation*}
K_{X}\left( f|g\right) =\left\langle \hat{X}\left( f\right) |\hat{X}\left(
g\right) \right\rangle _{\frak{h}}=\int_{\mathbb{\hat{R}}}f\left( \nu
\right) ^{\ast }g\left( \nu \right) \,dS_{X}\left( \nu \right) .
\end{equation*}

\subsection{Quantum Theory of WSS Processes}

A quantum probability space $\left( \frak{A},\mathbb{E}\right) $ consists of
a *-algebra $\frak{A}$\ of operators and a state $\mathbb{E}$\ on $\frak{A}$%
. A quantum stochastic process is then a family $X=\left( X_{t}\right) _{t}$
of elements of $\frak{A}$\ parameterized by time. Following the standard
interpretation of quantum mechanics, the self-adjoint elements of $\frak{A}$
correspond to the physical observables. More generally we consider
conditional expectations \cite{Davies}\cite{AC} where the partially-averaged
variables will live in some C*-subalgebra $\frak{B}$ of $\frak{A}$.

Let $\frak{B}$ be a C*-algebra and $\Phi :\frak{E}\mapsto \frak{B}$ be a
completely positive map. A kernel $\mathbb{K}:\frak{E}\times \frak{E}\mapsto 
\frak{B}$\ is defined by $\mathbb{K}\left( f|g\right) =\Phi \left( f^{\ast
}g\right) $. Then $\mathbb{C}\left( \tau \right) =\Phi \left( e_{\tau
}\right) $ is positive semi-definite on $\frak{B}$ in the sense that 
\begin{equation*}
\sum_{j,k=1}^{n}Z_{j}^{\ast }\mathbb{C}_{X}\left( t_{k}-t_{j}\right)
Z_{k}\geq 0
\end{equation*}
or all finite integers $n$, for arbitrary times $t_{1},\dots ,t_{n}\in 
\mathbb{R}$ and arbitrary operators $Z_{1},\dots ,Z_{n}\in \frak{B}$. This
suggests the following definition.

\begin{definition}
Let $\frak{A}$ be a C*-algebra, $\frak{B}$ a C*-subalgebra of $\frak{A}$ and 
$\mathbb{E}\left[ .|\frak{B}\right] $ a conditional expectation from $\frak{A%
}$ to $\frak{B}$. A mean-zero WSS quantum stochastic process $X=\left(
X_{t}\right) _{t}$ taking values in $\frak{A}$ and expected in $\frak{B}$ is
a family of operators $X_{t}\in \frak{A}$ such that $\mathbb{E}\left[ X_{t}|%
\frak{B}\right] =0$ and $\mathbb{C}_{X}\left( \tau \right) \mathbb{=E}\left[
X_{t}^{\ast }X_{t+\tau }|\frak{B}\right] $ is semi-positive definite on $%
\frak{B}$.
\end{definition}

In the situations where we are interested in a quantum probability space $%
\left( \frak{A},\mathbb{E}\right) $, we might typically ask that the
conditional expectation from $\frak{A}$ to $\frak{B}$ be compatible with the
state (i.e. $\mathbb{E}\left[ \mathbb{E}\left[ Z|\frak{B}\right] \right] =%
\mathbb{E}\left[ Z\right] $).

\bigskip

A natural example is where $H$ and $K$ are fixed Hilbert spaces and the
C*-algebras are concretely realized as $\frak{A}=\mathcal{B}\left( H\otimes
K\right) $, $\frak{B}=\mathcal{B}\left( H\right) $. The conditional
expectation then being partial expectation wrt. a fixed density matrix $\rho
_{K}$ on $K$: that is, 
\begin{equation*}
tr_{H}\left\{ T\;\mathbb{E}\left[ Z|\frak{B}\right] \right\} \equiv
tr_{H\otimes K}\left\{ T\otimes \rho _{K}\;Z\right\}
\end{equation*}
for all trace-class operators $T$ on $H$. The expectation $\mathbb{E}$\ will
be wrt. a product state $\rho =\rho _{H}\otimes \rho _{K}$.

\bigskip

Let $R$ be a Hilbert space and let $\mathcal{B}\left( H,R\right) $\ denote
the set of linear operators from $H$ to $R$. Suppose that we are given a map 
$V:\frak{E}\mapsto \mathcal{B}\left( H,R\right) ,$ then a semi-positive
definite kernel $\mathbb{K}:\frak{E}\times \frak{E}\mapsto \mathcal{B}\left(
H\right) $\ is obtained by setting $\mathbb{K}\left( f|g\right) =V\left(
f\right) ^{\ast }V\left( g\right) $. This is known as a Kolmogorov
decomposition of the kernel \cite{PS}. It is known that a kernel $\mathbb{K}$%
:$\frak{E}\times \frak{E}\mapsto \mathcal{B}\left( H\right) $\ admits a
Kolmogorov decomposition if and only if it is semi-positive definite. A
canonical choice is given by taking $R$ to be the closed linear manifold of $%
H$-valued functions on $\frak{E}$\ of the type $g\mapsto \mathbb{K}\left(
f|g\right) \xi $, where $f\in \frak{E}$ and $\xi \in H$, and taking $V\left(
g\right) \in \mathcal{B}\left( H,R\right) $ to be $V\left( f\right) :\xi
\mapsto \mathbb{K}\left( f|g\right) \xi $. The decomposition is minimal in
the sense that $R=\overline{\left\langle V\left( f\right) \xi :f\in \frak{E}%
,\xi \in H\right\rangle }$ and is unique up to unitary equivalence.

\bigskip

\begin{theorem}
Let $X=\left( X_{t}\right) _{t}$ be a mean-zero WSS quantum stochastic
process on $\left( \frak{A},\mathbb{E}\right) $ modelled on $\frak{B}$. Let $%
\frak{B}_{+}$ denote the cone of positive elements of $\frak{B}$. There
exists a non-decreasing right-continuous $\frak{B}_{+}$-valued function $%
\mathbb{S}_{X}$\ on $\mathbb{\hat{R}}$\ with $\lim_{\nu \rightarrow -\infty }%
\mathbb{S}_{X}\left( \nu \right) =0$ and $\lim_{\nu \rightarrow +\infty }%
\mathbb{S}_{X}\left( \nu \right) =\mathbb{C}_{X}\left( 0\right) $ such that 
\begin{equation*}
\mathbb{C}_{X}\left( t\right) =\int_{\mathbb{\hat{R}}}e^{2\pi i\,t\nu }\,d%
\mathbb{S}_{X}\left( \nu \right) .
\end{equation*}
$\mathbb{S}_{X}$ is called the spectral operator for $X$.
\end{theorem}

\bigskip

Now let $\frak{H}=L^{2}\left( \frak{A},\mathbb{E}\right) $\ be the vector
space of all operators $X\in \frak{A}$ such that $\mathbb{E}\left[ X^{\ast }X%
\right] <\infty $. Then $\frak{H}$ is a Hilbert space and if $X=\left(
X_{t}\right) _{t}$ is a quantum stochastic process on $\left( \frak{A},%
\mathbb{E}\right) $, then $\frak{H}_{X}=\overline{\left\langle X_{t}:t\in 
\mathbb{R}\right\rangle }$ will be a Hilbert subspace. We shall denote by $%
\frak{F}_{X}=L^{2}\left( \mathbb{\hat{R}},d\mathbb{S}_{X}\right) $ the space
of all measurable functions $f$ on $\mathbb{\hat{R}}$ such that $\int_{%
\mathbb{\hat{R}}}\left| f\left( \nu \right) \right| ^{2}\,d\mathbb{S}%
_{X}\left( \nu \right) $ defines a bounded element of $\frak{B}_{+}$.

\begin{theorem}
Let $X=\left( X_{t}\right) _{t}$ be a mean-zero WSS quantum stochastic
process on $\left( \frak{A},\mathbb{E}\right) $ modelled on $\frak{B}$ and
having spectral operator $\mathbb{S}_{X}$. Then there exists a linear map $%
\hat{X}:\frak{F}_{X}\mapsto \frak{H}_{X}$ such that 
\begin{equation*}
\mathbb{E}\left[ \hat{X}\left( f\right) ^{\ast }X\left( g\right) |\frak{B}%
\right] =\int_{\mathbb{\hat{R}}}f\left( \nu \right) ^{\ast }g\left( \nu
\right) \,d\mathbb{S}_{X}\left( \nu \right) .
\end{equation*}
\end{theorem}

\bigskip

In analogy to the classical case, we denote by $\hat{X}\left[ d\nu \right] $
the corresponding $\frak{B}_{+}$-valued measure on $\mathbb{\hat{R}}$.

\section{Linear Quantum Filters}

The notion of a linear filtering is of considerable importance in stochastic
modelling \cite{A}\cite{T}. It is possible to extend this to the quantum
case.

\bigskip

\begin{definition}
Let $\psi _{L}:\mathbb{\hat{R}}\mapsto \frak{B}$ be measurable. A linear
filter $L$ with characteristic function $\psi _{L}\left( \nu \right) $
acting on the $\frak{A}$-valued mean-zero WSS quantum stochastic processes
expected on $\frak{B}$ is the linear transformation defined by $X\mapsto LX$
where 
\begin{equation*}
\left( LX\right) _{t}:=\int_{\mathbb{\hat{R}}}e^{2\pi i\;\nu t}\,\hat{X}%
\left[ d\nu \right] \,\psi _{L}\left( \nu \right)
\end{equation*}
and the domain of $L$ is the set $\mathcal{D}_{L}$ of all such processes $X$
for which the integral $\int_{\mathbb{\hat{R}}}\psi _{L}\left( \nu \right)
^{\ast }\,d\mathbb{S}_{X}\left( \nu \right) \,\psi _{L}\left( \nu \right) $
converges in $\frak{B}$.
\end{definition}

Given $X\in \mathcal{D}_{L}$, the process $LX$ will again be mean-zero WSS
process and its spectral measure will be $d\mathbb{S}_{LX}\left( \nu \right)
=\psi _{L}\left( \nu \right) ^{\ast }\,d\mathbb{S}_{X}\left( \nu \right)
\,\psi _{L}\left( \nu \right) $.

\bigskip

\subsection{Examples}

We begin with some simple ``classical'' filters having $c$-number
characteristic functions:

\begin{itemize}
\item  Time shifts $\left( LX\right) _{t}=X_{t+s};$ $\psi _{L}=e_{s}.$

\item  Time derivatives $\left( LX\right) _{t}=\dot{X}_{t};$ $\psi
_{L}\left( \nu \right) =2\pi i\nu $

\item  Scalar convolutions $\left( LX\right) _{t}=\left( X\ast h\right)
_{t}; $ $\psi _{L}\left( \nu \right) =\hat{h}\left( \nu \right) $
\end{itemize}

\bigskip

However, filters with operator-valued characteristic functions are possible.
Let $A\in \frak{B}$ and $\Gamma \in \frak{B}_{+}$; then set $h\left(
t\right) =e^{-\Gamma t}A\theta \left( t\right) $. We can consider a process $%
X$ passed through the linear filter corresponding to convolution with $h$.
The output process $Y$ will have spectral density $d\mathbb{S}_{Y}\left( \nu
\right) =A^{\ast }\frac{1}{2\pi i-\Gamma }\,d\mathbb{S}_{X}\left( \nu
\right) \,\frac{1}{2\pi i+\Gamma }A$. As a specific case, we could consider $%
X$ to be a white noise input with spectrum $d\mathbb{S}_{X}\left( \nu
\right) =Sd\nu $ where $S\in \frak{B}_{+}$. (For instance, $X$ might be
combinations of creator or annihilator white noise \cite{Gard}\cite{H}\cite
{Obata}.) Then provided the damping operator $\Gamma $ commutes with $S$,
the output will have a continuous Lorentzian-type spectral density, no
singular or pure point component, and its covariance will be the
Ornstein-Uhenbeck type $\mathbb{C}_{X}\left( t\right) =A^{\ast }Se^{-\Gamma
\left| t\right| }A$. If $\left[ \Gamma ,S\right] \neq 0$ then the spectrum
will be more complicated.

\end{document}